\documentclass[prb,twocolumn,showpacs,preprintnumbers,amsmath,amssymb]{revtex4}

\usepackage{graphicx}
\usepackage{dcolumn}
\usepackage{bm}

\begin{document}

\preprint{APS/123-QED}

\title{Pressure-induced superconductivity in Eu$_{0.5}$Ca$_{0.5}$Fe$_2$As$_2$ : \\ FeAs-based superconductivity hidden by antiferromagnetism of Eu sublattice
}

\author{Akihiro Mitsuda}\email{3da@phys.kyushu-u.ac.jp}
\author{Tomohiro Matoba}%
\author{Hirofumi Wada}%
\affiliation{Department of Physics, Kyushu University, 6-10-1 Hakozaki, Higashi-ku, Fukuoka 812-8581, Japan}
\author{Fumihiro Ishikawa}
\affiliation{%
Graduate School of Science and Technology, Niigata University, 
Niigata 950-2181, Japan
}%
\author{Yuh Yamada}
\affiliation{%
Department of Physics, Niigata University, Niigata 950-2181, Japan}%

\date{\today}

\begin{abstract}
To clarify superconductivity in EuFe$_2$As$_2$ hidden by antiferromagnetism 
of Eu$^{2+}$, we investigated a Ca-substituted sample,
Eu$_{0.5}$Ca$_{0.5}$Fe$_2$As$_2$, under high pressure.
For ambient pressure, the sample exhibits a spin-density-wave (SDW) 
transition at $T_\mathrm{SDW}=191$~K 
and antiferromagnetic order at $T_\mathrm{N}=4$~K, but no evidence of
superconductivity down to 2~K.
The Ca-substitution certainly weakens the antiferromagnetism.
With increasing pressure, $T_\mathrm{SDW}$ shifts to lower temperature
and becomes more unclear. Above 1.27~GPa, pressure-induced superconductivity
with zero resistivity is observed at around $T_\mathrm{c}=20$~K. At 2.14~GPa,
$T_\mathrm{c}$ reaches a maximum value of 24~K and the superconducting
transition becomes the sharpest. 
These features of emergence of the superconductivity are qualitatively 
similar to
those observed in $A$Fe$_2$As$_2$ ($A=\mathrm{Ba}$, Ca).

\end{abstract}

\pacs{74.62.Fj, 74.62.Dh, 74.25.Dw, 74.70.Dd}
\maketitle

Since Kamihara et al. found that LaFeAsO$_{1-x}$F$_x$ for 
$0.05\le x\le 0.12$ is a high-$T_\mathrm{c}$ superconductor of 
$T_\mathrm{c}=26$~K,\cite{kamihara} a lot of active researches on Fe-based 
superconductors have been started all over the world.
Presently, replacing La with other lanthanides raised $T_\mathrm{c}$
up to 55~K, which is the highest $T_\mathrm{c}$ value 
except the high-$T_\mathrm{c}$ cuprates.\cite{SmFeAsO}
The parent material LaFeAsO crystallizes in the tetragonal ZrCuSiAs type
structure, in which LaO and FeAs layers are stacked alternately, and performs
a structural phase transition from tetragonal to orthorhombic symmetry
at 155~K\cite{cruz,nomura} and a spin-density-wave (SDW) transition 
at 137~K.\cite{cruz}
Either F-doping or oxygen vacancies, which corresponds to electron doping, 
suppresses these transitions and induces superconductivity.\cite{kamihara}
Though Fe is a magnetic element, which has been considered as destructive
to superconductivity, it is intriguing that the FeAs layer plays a key 
role to giving rise to the superconductivity. 

Through the active researches, similar superconductivity of 
$T_\mathrm{c}=38$~K was found also in Ba$_{1-x}$K$_x$Fe$_2$As$_2$.\cite{rotter}
This system forms the tetragonal ThCr$_2$Si$_2$ type structure, 
which consists of alternate stacking of a Ba/K layer and similar FeAs layer 
to that in LaFeAsO. 
Similarly to LaFeAsO, the parent compound 
BaFe$_2$As$_2$ exhibits the SDW transition at 140~K.\cite{rotter,rotter2}
Substitution of K for Ba,
which corresponds to hole doping, induces the superconductivity,
accompanied by collapse of the SDW transition.\cite{rotter}
In addition, applying external pressure \cite{ishikawa} and/or 
chemical pressure \cite{jiang,kasahara} (isovalent substitution of P for As) 
also induces the superconductivity.
EuFe$_2$As$_2$ with the same structure as BaFe$_2$As$_2$ exhibits 
the SDW transition due to Fe at 190~K and antiferromagnetic order of Eu$^{2+}$ 
at 20~K.~\cite{ren}
With increasing pressure, the SDW transition is continuously 
collapsed and a sharp drop at 30~K in electrical resistivity appears 
at around 2~GPa, which is reminiscent of superconductivity. 
After the drop, however, the resistivity goes to a finite value 
(not zero resistivity) at around 20~K, where the antiferromagnetic order 
occurs.\cite{miclea} 
It is speculated that cooper pairs should be destroyed by 
molecular field due to antiferromagnetism (AFM) of Eu$^{2+}$ ions
as is observed in HoMo$_6$S$_8$ \cite{mishikawa} an 
HoNi$_2$B$_2$C.\cite{eisaki}
To clarify the superconducting properties hidden by the antiferromagnetic 
order of Eu$^{2+}$, 
we try to weaken the molecular field by 50\% diluting Eu$^{2+}$ by 
nonmagnetic isovalent Ca$^{2+}$, which corresponds to decrease in number 
of magnetic moments of Eu$^{2+}$ without carrier doping. 
Additionally, the diluted sample was investigated under high pressure.
In the present study, we report pressure-induced superconductivity and
zero resistivity in Eu$_{0.5}$Ca$_{0.5}$Fe$_2$As$_2$. 
It is shown that Eu ions enhance
superconducting transition temperature $T_\mathrm{c}$.

The single crystal of Eu$_{0.5}$Ca$_{0.5}$Fe$_2$As$_2$ was grown 
in a tin flux. A mixture of constituent elements in the ratio of 
$\mathrm{Eu}:\mathrm{Ca}:\mathrm{Fe}:\mathrm{As}:\mathrm{Sn}=0.5:0.5:2:2:48$, 
which was put into an alumina crucible, 
was heated at 1000$^\circ$C for 24 hours in a quartz tube,
where 1/3-atm Ar gas at room temperature was sealed.
Subsequently, the mixture was cooled down to 500$^\circ$C at the rate 
of $-14^\circ\mathrm{C/h}$ and the tin flux was removed by a centrifuge.
We obtained many pieces of plate-like single crystals with 
typical dimensions of $\sim 3\times 3\times 0.1$~mm$^3$.
Powder x-ray diffraction pattern at room temperature exhibits 
a single phase which crystallizes in the tetragonal ThCr$_2$Si$_2$ type
structure with lattice constants of $a=3.897$~\AA\ and $c=12.006$~\AA.
These values lie almost halfway between 
EuFe$_2$As$_2$ ($a=3.902$~\AA\ and $c=12.138$~\AA) and
CaFe$_2$As$_2$ ($a=3.886$~\AA\ and $c=11.776$~\AA), which 
are also synthesized by our group. The lattice constants of both end compounds
are in good agreement with those reported in Ref.~\onlinecite{park,ren}.
The Energy dispersive X-ray (EDX) spectroscopy confirms that 
composition of the sample is almost
$\mathrm{Eu}:\mathrm{Ca}:\mathrm{Fe}:\mathrm{As}=0.5:0.5:2:2$.
These results suggests 50\% Eu atoms are substituted by Ca atoms homogeneously.
Electrical resistivity under high pressure was measured with current
flowing in the $ab$ plane by using an ac resistance bridge 
(LR-700, Linear Research) in the temperature range between 4.2 and 280~K. 
Pressure was generated up to 2.47~GPa by using a piston-cylinder type
pressure cell, which consists of inner (NiCrAl alloy) and outer (CuBe alloy) 
cylinders. The sample and a tin manometer were placed into a Teflon cell
filled with a pressure transmitting medium of mixture of 2 types of 
Flourinert in the ratio of $\mathrm{FC70}:\mathrm{FC77}=1:1$. 
The Teflon cell was inserted into the pressure cell and pressed by 
pistons made of nonmagnetic tungsten carbide.
Magnetization measurement under high pressure was carried out by 
a SQUID magnetometer (MPMS, Quantum Design)
in the temperature range between 2 and 300~K. Pressure was generated
up to 0.8~GPa by basically the same method as in the resistivity measurement.
We used a pressure cell made of nonmagnetic CuTi alloy and pistons 
made of zirconia to reduce a magnetization signal from the pressure cell. 
Many pieces of small single crystals directed randomly were used 
in the magnetization measurement. To obtain magnetization of
the sample, magnetization of the pressure cell was subtracted
from the measured value.
 

\begin{figure}
\includegraphics[width=7cm]{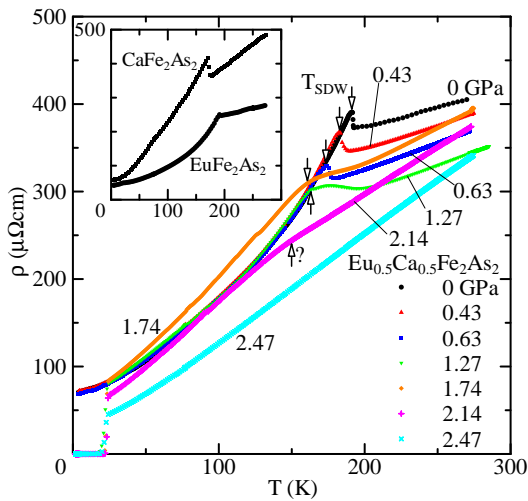}
\caption{\label{fig:rhovsT} Electrical resistivity versus temperature 
of Eu$_{0.5}$Ca$_{0.5}$Fe$_2$As$_2$ at various pressures. The open 
arrows depict spin-density-wave (SDW) temperature, $T_\mathrm{SDW}$. 
However, $T_\mathrm{SDW}$ is difficult to determine at 2.14~GPa.
The inset exhibits temperature dependence of the resistivity 
of CaFe$_2$As$_2$ and EuFe$_2$As$_2$ synthesized by our group.}
\end{figure}

\begin{figure}
\includegraphics[width=7cm]{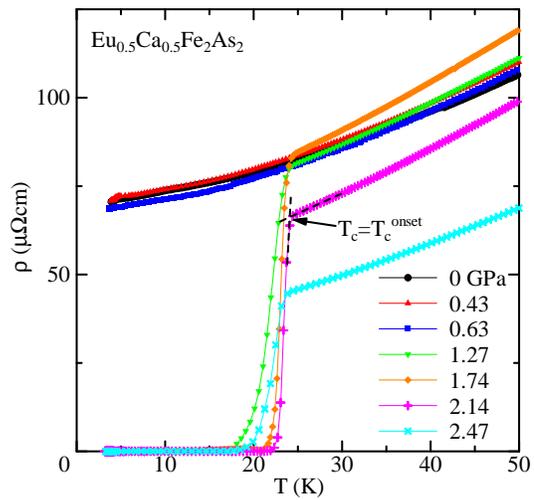}
\caption{\label{fig:rhovsT_LT} Electrical resistivity versus temperature 
of Eu$_{0.5}$Ca$_{0.5}$Fe$_2$As$_2$ at various pressures in the low
temperature region. The short dash line depicts how to define onset
superconducting temperature $T_\mathrm{c}^\mathrm{onset}$.}
\end{figure}

Figure~\ref{fig:rhovsT} demonstrates temperature dependence of 
electrical resistivity $\rho$ of Eu$_{0.5}$Ca$_{0.5}$Fe$_2$As$_2$ 
under various pressures. 
In addition, the
$\rho$--$T$ curves of CaFe$_2$As$_2$ and EuFe$_2$As$_2$ synthesized by 
our group, of which $T_\mathrm{SDW}$ are 170~K and 191~K, respectively, 
are also shown in the inset. These curves qualitatively coincide
with those reported in Ref.~\onlinecite{park,ni2,ren}.
The behavior of Eu$_{0.5}$Ca$_{0.5}$Fe$_2$As$_2$ for 
ambient pressure is intermediate between both end compounds.
A sharp cusp, which corresponds to spin-density-wave
(SDW) transition, is observed at $T_\mathrm{SDW}=191$~K. $T_\mathrm{SDW}$
is closed to that of EuFe$_2$As$_2$,\cite{ren} but shape of the cusp is 
similar to that of CaFe$_2$As$_2$.\cite{ni2}
Below $T_\mathrm{SDW}$, with decreasing 
temperature, the resistivity decreases significantly with concave-up curvature
down to $70\,\mu\Omega\mathrm{cm}$ without any anomalies like 
superconductivity or AFM of Eu$^{2+}$. 
Unlike Eu$_{0.5}$K$_{0.5}$Fe$_2$As$_2$,\cite{jeevan} substitution of 
isovalent Ca for Eu neither collapses the SDW transition nor 
induces superconductivity.
At 0.43 and 0.63~GPa, except for $T_\mathrm{SDW}$, which shifts to lower
temperatures with increasing pressure, 
temperature dependence of $\rho$ is qualitatively the same 
as that at ambient pressure.
At 1.27~GPa, superconductivity suddenly appears at around 23~K,
which is much higher than that of CaFe$_2$As$_2$.
The cusp of the SDW transition transforms into a broad maximum. 
The $\rho$--$T$ curve
is comparably similar to that below 0.63~GPa except for the superconducting
behavior and the shape of $T_\mathrm{SDW}$, which exhibits this pressure is 
critical 
pressure $P_\mathrm{c}$.
At 1.74~GPa, complete zero resistivity is 
realized and the superconducting transition becomes sharper than that 
at 1.27~GPa, which suggests more homogeneous superconductivity. 
The anomaly associated with the SDW transition becomes broader and
more indistinct. 
The slope of the $\rho$--$T$ curve above $T_\mathrm{SDW}$ becomes larger.
These features result in considerable change of the shape of the 
$\rho$--$T$ curve.
At 2.14~GPa, the sharp superconducting transition shifts
to a little higher temperature, which demonstrates maximum $T_\mathrm{c}$
of 24~K in the present study. The anomaly of the SDW transition can 
be seen as slight
convex-up behavior at around 150~K, but is too shallow to determine 
$T_\mathrm{SDW}$ precisely.
The residual resistivity $\rho_0$, which is determined as the $\rho$ value 
just above $T_\mathrm{c}$, begins to drop.
Finally, at 2.47~GPa, again the transition  becomes less sharp and 
shift to lower temperature, which is possibly onset of collapse of the
superconductivity as is observed in CaFe$_2$As$_2$~\cite{torikachivili,lee} 
and BaFe$_2$As$_2$.\cite{ishikawa} 
The $\rho$--$T$ curve above $T_\mathrm{c}$ exhibits 
concave up behavior and no anomaly, which indicates the SDW transition 
is collapsed completely. The $\rho_0$ value decreases down to 
$\sim 45\,\mu\Omega\mathrm{cm}$. Similar drop of $\rho_0$ with pressure
is observed also in CaFe$_2$As$_2$.\cite{torikachivili,lee}

\begin{figure}
\includegraphics[width=7cm]{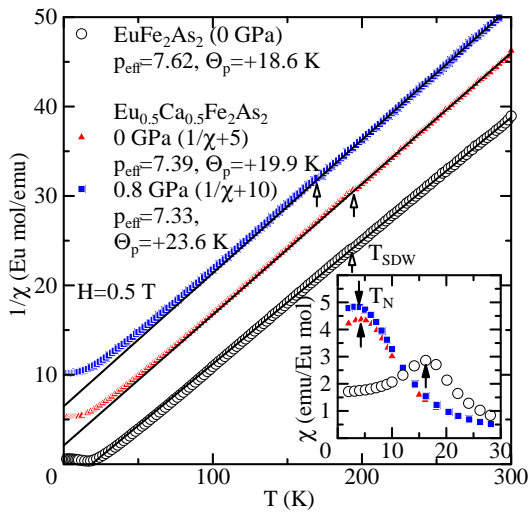}
\caption{\label{fig:isusvsT} Inverse susceptibility versus temperature 
of EuFe$_2$As$_2$ at 0~GPa and of Eu$_{0.5}$Ca$_{0.5}$Fe$_2$As$_2$ 
at 0 and 0.8~GPa. 
The open arrows depict spin-density-wave (SDW) temperature. 
The solid lines exhibits Curie-Weiss law.
The inset demonstrates
temperature dependence of magnetic susceptibility in the low temperature
region. The solid arrows show the N\'eel temperature.}
\end{figure}

Figure~\ref{fig:isusvsT}
shows temperature dependence of inverse magnetic susceptibility $1/\chi$
of EuFe$_2$As$_2$ at 0~GPa and of Eu$_{0.5}$Ca$_{0.5}$Fe$_2$As$_2$ 
at 0 and 0.8~GPa. The inset indicates the $\chi$--$T$ curves in the low
temperature region.
For EuFe$_2$As$_2$, the data is almost in accordance with the previous 
one.~\cite{ren}
There exist a clear peak in the $\chi$--$T$ curve at $T_\mathrm{N}=17$~K
and a kink in the $1/\chi$--$T$ curve at $T_\mathrm{SDW}=193$~K.
Above $T_\mathrm{SDW}$, the susceptibility obeys the Curie-Weiss (CW)
law with an effective moment $p_\mathrm{eff}$ of 
$7.62\mu_\mathrm{B}/\mathrm{Eu}$ and 
a Weiss temperature $\Theta_\mathrm{p}$ of $+18.6$~K.
The $p_\mathrm{eff}$ value is close to the theoretical value 
($7.94\mu_\mathrm{B}/\mathrm{Eu}$) of a free Eu$^{2+}$ ion. The positive
$\Theta_\mathrm{p}$ value suggests a ferromagnetic interaction.
Actually, a comparably small magnetic field of $\sim 1$~T
can saturate magnetization.\cite{ren}
Substituting Eu with Ca by 50~\% lowers $T_\mathrm{N}$ down to 4~K, 
but nearly retains the CW behavior. The former means
that the AFM is destabilized, which results from depression
of RKKY (Ruderman-Kasuya-Kittel-Yosida) interaction due to dilution of
magnetic Eu$^{2+}$ ions. The latter implies that the Eu valence remains
almost $2+$.
With applying pressure to Eu$_{0.5}$Ca$_{0.5}$Fe$_2$As$_2$, 
$T_\mathrm{SDW}$ is lowered, as shown also in Fig.~\ref{fig:rhovsT}, 
but the $T_\mathrm{N}$ value and the CW behavior, which are associated
with Eu, are nearly unchanged.
This means that the valence and the magnetism of Eu are insensitive to 
pressure.
We also measured magnetization curves at 1.8~K. (not shown) 
At both 0 and 0.8~GPa, the magnetization is saturated to 
$6.5\mu_\mathrm{B}/\mathrm{Eu}$, which is close to 
magnetic moment of Eu$^{2+}$ of $7\mu_\mathrm{B}/\mathrm{Eu}$.
This also supports Eu valence remains close to $2+$ under pressures up to 
0.8~GPa. We speculate that the magnetism and the valence of Eu are probably 
preserved also under higher pressures than 0.8~GPa. 
To investigate competition between the magnetism of Eu and the 
superconductivity in more detail, 
magnetization measurements under higher pressure
are now in progress.

\begin{figure}
\includegraphics[width=7cm]{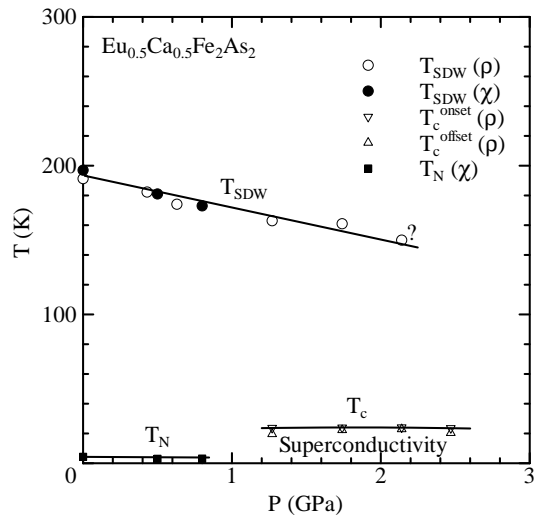}
\caption{\label{fig:TPphasediagram} Temperature versus pressure phase diagram 
of Eu$_{0.5}$Ca$_{0.5}$Fe$_2$As$_2$. The $T_\mathrm{c}^\mathrm{onset}$ 
value has a faint maximum of 24~K at around 2.14~GPa.}
\end{figure}

To summarize the present experiments, the $T$--$P$ phase diagram
is shown in Fig.~\ref{fig:TPphasediagram}. We define 
$T^\mathrm{offset}_\mathrm{c}$ as temperature where the resistivity 
reaches 10\% of the residual resistivity $\rho_0$. 
As $T_\mathrm{SDW}$ is being 
suppressed with increasing pressure, the superconductivity of 
$T_\mathrm{c}\sim 20$~K emerges suddenly at around 1.2~GPa. 
Pressure dependence of $T_\mathrm{c}$ is quite small but 
has a faint maximum of 24~K at around 2.14~GPa, where the 
superconducting transition is the sharpest. The AFM of Eu$^{2+}$ is 
realized at around 4~K. Since $T_\mathrm{N}$ is much lower than $T_\mathrm{c}$
unlike EuFe$_2$As$_2$, it seems that the superconductivity is realized stably
in spite of the AFM. 

Compared with CaFe$_2$As$_2$ ($P_\mathrm{c}\sim0.4$~GPa and $T_\mathrm{c}
\sim 10$~K),\cite{torikachivili,lee} 
both $P_\mathrm{c}$ and $T_\mathrm{c}$ are quite higher in Eu$_{0.5}$Ca$_{0.5}$Fe$_2$As$_2$
($P_\mathrm{c}\sim1.3$~GPa and $T_\mathrm{c}\sim 24$~K),
which indicates that the superconductivity observed in this study is not
due to an impurity of CaFe$_2$As$_2$ but intrinsic.
In addition, it is suggested that Eu ions play a key role in enhancing 
$T_\mathrm{c}$. Though a structural phase transition from tetragonal
to collapsed tetragonal type takes place above 0.8~GPa in CaFe$_2$As$_2$,
there exists no evidence of such a structural transition under 
high pressures up to 2.47~GPa in Eu$_{0.5}$Ca$_{0.5}$Fe$_2$As$_2$.
On the other hand, 
compared with EuFe$_2$As$_2$ ($P_\mathrm{c}\sim 2.0$~GPa and $T_\mathrm{c}
\sim 30$~K, but no zero resistivity due to AFM of Eu$^{2+}$),\cite{miclea} 
both $P_\mathrm{c}$ and $T_\mathrm{c}$ are lower in the present system.
Since lattice volume of Eu$_{0.5}$Ca$_{0.5}$Fe$_2$As$_2$ is smaller 
by 1.4\% than that of EuFe$_2$As$_2$, it can be regarded that chemical
pressure is applied in addition to hydrostatic pressure in the present 
study. As a result, $P_\mathrm{c}$ in Eu$_{0.5}$Ca$_{0.5}$Fe$_2$As$_2$ is 
smaller than that in EuFe$_2$As$_2$ ($\sim 2.0$~GPa).\cite{miclea} 
Eu ions has two opposite effects. One is to induce
superconductivity of $T_\mathrm{c}\sim 30$~K. The other is to
destroy cooper pairs through the AFM. Dilution of Eu by Ca weakens 
both effects, which results in decrease in $T_\mathrm{c}$ and appearance
of zero resistivity. 
Recently Terashima et al. have reported that EuFe$_2$As$_2$
becomes a superconductor with zero resistivity under high pressure of 
2.8~GPa.\cite{terashima}
Matsubayashi et al. also observe zero resistivity in EuFe$_2$As$_2$
in the quite narrow pressure range of $2.7\sim 2.8$~GPa.\cite{matsubayashi} 
In the limited pressure
range of $\sim 2.8$~GPa, the superconductivity might slightly overcome the AFM.
Very recently, Zheng et al. demonstrate that both diluting Eu with Sr or Ba
and substitution of Co for Fe induces superconductivity 
with zero resistivity,\cite{zheng}
which coincides with our results.
The chemical and hydrostatic
pressures are also expected to induce valence transition from Eu$^{2+}$
toward Eu$^{3+}$ because Eu$^{2+}$ has larger volume than Eu$^{3+}$ does.
However, the effective magnetic moment and the spontaneous magnetization 
remain the values for a Eu$^{2+}$ ion under pressures up to 0.8~GPa 
in the present system. There exists no evidence of
Eu valence change in the present study.

In conclusion, we observed pressure-induced superconductivity 
($T_\mathrm{c}\sim 24$~K) with zero resistivity 
of Eu$_{0.5}$Ca$_{0.5}$Fe$_2$As$_2$ in the pressure range of 
$P=1.27\sim 2.47$~GPa. Diluting Eu with isovalent Ca weakens the 
AFM of Eu$^{2+}$ with preserving the SDW and the CW behavior. 
It is strongly suggested that in EuFe$_2$As$_2$ the AFM of Eu$^{2+}$ and 
the pressure-induced superconducitivity compete with each other.
Substitution of Ca and/or applying pressure seem not to change 
Eu valence.

\begin{acknowledgments}
The authors thank Dr. M. Watanabe at Center of Advanced Instrumental
Analysis, Kyushu University for helping us to perform SEM-EDX 
analysis.
\end{acknowledgments}


\begin{thebibliography}{99}
\bibitem{kamihara}
Y. Kamihara, T. Watanabe, M. Hirano, and H. Hosono, J. Am. Chem. Soc. 
{\bf 130}, 3296 (2008).

\bibitem{SmFeAsO}
X. F. Chen, T. Wu, G. Wu, R. H. Liu, H. Chen, and D.
F. Fang, Nature (London) {\bf 453}, 761 (2008).

\bibitem{cruz}
C. de la Cruz, Q. Huang, J. W. Lynn, J. Li, W. Ratcliff II, H. A.
Mook, G. F. Chen, J. L. Luo, N. L. Wang, and Pengcheng Dai,
Nature (London) {\bf 453}, 899(2008).

\bibitem{nomura}
T. Nomura, S.W. Kim, Y. Kamihara, M. Hirano, P. V. Sushko, K.
Kato, M. Takata, A. L. Shluger, and H. Hosono, Supercond. Sci. Technol.
{\bf 21}, 125028 (2008).


\bibitem{rotter}
M. Rotter, M. Tegel, and D. Johrendt,
Phys. Rev. Lett. {\bf 101}, 107006 (2008).


\bibitem{rotter2}
M. Rotter, M. Tegel, and D. Johrendt,
Phys. Rev. B {\bf 78}, 020503 (2008).

\bibitem{ishikawa}
F. Ishikawa, N. Eguchi, M. Kodama, K. Fujimaki, M. Einaga, A. Ohmura, 
A. Nakayama, A. Mitsuda, and Y. Yamada, Phys. Rev. B {\bf 79}, 172506 (2009).

\bibitem{jiang}
S. Jiang, H. Xing, G. Xuan, C. Wang, Z. Ren,
C. Feng, J. Dai, Z. Xu, and G. Cao, J. Phys.: Condens. Matter {\bf 21}, 
382203 (2009).

\bibitem{kasahara}
S. Kasahara, T. Shibauchi, K. Hashimoto, K. Ikada,
S. Tonegawa, H. Ikeda, H. Takeya, K. Hirata,
T. Terashima, and Y. Matsuda, arXiv:0905.4427 (2009).

\bibitem{ren}
Z. Ren, Z. Zhu, S. Jiang, X. Xu, Q. Tao, C. Wang, 
C. Feng, G. Cao, and Z. Xu, Phys. Rev. B {\bf 78}, 052501 (2008).
\bibitem{miclea}
C. F. Miclea, M. Nicklas, H. S. Jeevan, D. Kasinathan, Z. Hossain, H. Rosner, 
P. Gegenwart, C. Geibel, and F. Steglich, Phys. Rev. B {\bf 79}, 212509 (2009).

\bibitem{mishikawa}
M. Ishikawa and \O. Fischer, Solid State Commun. {\bf 23}, 37 (1977).

\bibitem{eisaki}
H. Eisaki, H. Takagi, R. J. Cava, B. Batlogg, J. J. Krajewski, W.
F. Peck, Jr., K. Mizuhashi, J. O. Lee, and S. Uchida, Phys. Rev.
B {\bf 50}, 647 (1994).

\bibitem{park}
T. Park, E. Park, H. Lee, T. Klimczuk, E. D. Bauer, F. Ronning, 
and J. D. Thompson, J. Phys. condes. Matter {\bf 20} 322204 (2008).

\bibitem{ni2}
N. Ni, S. Nandi, A. Kreyssig, A. I. Goldman, E. D. Mun, S. L. Bud'ko, 
and P. C. Canfield, Phys. Rev. B {\bf 78}, 014523 (2008).

\bibitem{jeevan}
H. S. Jeevan, Z. Hossain, Deepa Kasinathan, H. Rosner, C. Geibel, and 
P. Gegenwart, Phys. Rev. B {\bf 78} 092406 (2008).

\bibitem{torikachivili}
M. S. Torikachvili, S. L. Bud'ko, N. Ni, and P. C. Canfield, 
Phys. Rev. Lett. {\bf 101}, 057006 (2008).

\bibitem{lee}
H. Lee, E. Park, T. Park, V. A. Sidorov, F. Ronning, E. D. Bauer, 
and J. D. Thompson, Phys. Rev. B {\bf 80}, 024519 (2009).



\bibitem{terashima}
T. Terashima, M. Kimata, H. Satsukawa, A. Harada, K. Hazama, S. Uji,
H. S. Suzuki, T. Matsumoto, and K. Murata, J. Phys. Soc. Jpn. {\bf 78}, 083701 
(2009).

\bibitem{matsubayashi}
K. Matsubayashi, in preparation

\bibitem{zheng}
Q. J. Zheng, Y. He, T. Wu, G. Wu, H. Chen, J. J. Ying, R. H. Liu,
X. F. Wang, Y. L. Xie, Y. J. Yan, Q. J. Li, and X. H. Chen, 
arXiv:0907.5547v1.

\end{thebibliography}
\end{document}